\documentclass[prd,referee,nopacs,nofootinbib, twocolumn]{revtex4}
\usepackage{amsfonts}
\usepackage{amsmath}
\usepackage{eurosym}
\usepackage{amssymb}
\usepackage{graphicx}
\usepackage{color}

\newcommand{\f}[2]{\frac{#1}{#2}}
\def\be{\begin{equation}}
\def\ee{\end{equation}}
\def\bea{\begin{eqnarray}}
\def\eea{\end{eqnarray}}

\begin{document}

\title{The Einstein dark energy model}
\author{Zahra Haghani}
\email{z.haghani@du.ac.ir}
\affiliation{School of Physics, Damghan University, Damghan, Iran,}
\author{Tiberiu Harko}
\email{t.harko@ucl.ac.uk}
\affiliation{ Department of Physics, Babes-Bolyai University, Kogalniceanu Street,
Cluj-Napoca 400084, Romania,}
\affiliation{School of Physics, Sun Yat-Sen University, Guangzhou 510275, People's
Republic of China}
\affiliation{Department of Mathematics, University College London, Gower Street,
London WC1E 6BT, United Kingdom}
\author{Shahab Shahidi$^1$}
\email{s.shahidi@du.ac.ir}
\affiliation{School of Physics, Damghan University, Damghan, Iran}

\date{\today }

\begin{abstract}
In 1919 Einstein tried to solve the problem of the structure of matter by
assuming that the elementary particles are held together solely by
gravitational forces. In addition, Einstein also assumed the presence inside
matter of electromagnetic interactions.
Einstein showed that the cosmological constant can be
interpreted as an integration constant, and that the energy content of the
Universe should consist of 25\% gravitational energy, and 75\%
electromagnetic energy. In the present paper we reinterpret Einstein's
elementary particle theory as a vector type dark energy model, by assuming
a gravitational action containing a linear combination of the Ricci
scalar and the trace of the matter energy-momentum tensor, as well as a
massive self-interacting vector type dark energy field, coupled with the
matter current. Since in this model the matter energy-momentum tensor is not
conserved, we interpret these equations from the point of view of the
thermodynamics of open systems as describing matter creation from the
gravitational field. In the vacuum case the model admits a de Sitter type
solution. The cosmological parameters, including Hubble function, deceleration parameter, matter energy density are obtained as a function of the redshift by using analytical and numerical techniques, and for different values of the  model parameters. For all considered cases the Universe experiences an accelerating expansion, ending with a de Sitter type evolution.
\end{abstract}

\maketitle

\tableofcontents

\section{Introduction}

After proposing a static model of the Universe, based on the introduction of
the cosmological constant in the gravitational field equations \cite{Ein1},
Einstein tried to solve the problem of the structure of the elementary
particles \cite{Ein2}. Adopting the basic assumption that material particles
are bound together only by the gravitational force, depending on the metric
tensor $g_{\mu \nu}$ and its derivatives, Einstein considered that the
mass-energy structure of the elementary particles is described by an
energy-momentum tensor of the Maxwell type, $S_{\mu \nu}$, constructed from
the electromagnetic fields $F_{\mu \nu}$, which should be proportional with
a differential expression of second order formed from the metric tensor
coefficients (a linear combination of the Ricci and metric tensors). Hence Einstein proposed as the basic equation describing the
properties of elementary particles the equation \cite{Ein2}
\begin{equation}  \label{1}
R_{\mu \nu}+\bar{\lambda}g_{\mu \nu }R=\kappa ^2 S_{\mu \nu},
\end{equation}
where $\bar{\lambda}$ is a constant, and $\kappa ^2 =8\pi G/c^4$ is the
gravitational coupling constant. Since $S_{\mu}^{\mu}=0$, $\bar{\lambda}$
can be determined by taking the trace of Eq.~(\ref{1}) as $\bar{\lambda}%
=-1/4 $, and thus Einstein's equation (\ref{1}) becomes
\begin{equation}  \label{2}
R_{\mu \nu}-\frac{1}{4}g_{\mu \nu }R=\kappa ^2 S_{\mu \nu}.
\end{equation}
Moreover, Einstein assumed that the electromagnetic field satisfy Maxwell's
equations. Thus, by taking the covariant derivative of Eq.~(\ref{2}) we
obtain
\begin{equation}  \label{3}
\frac{1}{4}\nabla _{\mu}R=-\frac{\kappa ^2 }{c}F_{\mu \nu}j^{\nu},
\end{equation}
where $j^{\nu}$ is the electric current. Eq.~(\ref{3}) shows that the
Coulomb repulsive forces are held in equilibrium by a gravitational
pressure, thus assuring the stability of elementary particles. In the vacuum
outside the elementary particle, and in the absence of charges, Eq.~(\ref{3}%
) gives
\begin{equation}  \label{3a}
R=R_0=\mathrm{constant}.
\end{equation}
At this moment Einstein did proceed to the determination of the matter
energy-momentum tensor. He assumed that the gravitational field equations
containing the cosmological constant,
\begin{equation}  \label{4}
R_{\mu \nu}-\frac{1}{2}g_{\mu \nu}R+\Lambda g_{\mu \nu}=\kappa ^2 T_{\mu
\nu},
\end{equation}
also remain valid. For the free space, in the absence of matter, Eq.~(\ref{4}%
) gives $R=4\Lambda$, and by comparison with Eq.~(\ref{3a}), it follows that
$\Lambda =R_0/4$. This result represents one of the main advantages of this
approach, since it shows that \textit{the cosmological constant may not be a
characteristic of the fundamental law of gravity}, but an \textit{%
integration constant}. Hence Eq.~(\ref{4}) becomes
\begin{equation}
G_{\mu \nu}+\frac{R_0}{4}g_{\mu \nu}=\kappa ^2 T_{\mu \nu},
\end{equation}
giving
\begin{equation}
\kappa ^2 T=\left(R_0-R\right),
\end{equation}
while Eq.~(\ref{2}) can be reformulated as
\begin{equation}
G_{\mu \nu}+\frac{1}{4}Rg_{\mu \nu}=\kappa ^2 S_{\mu \nu}.
\end{equation}
Comparison of the above equations shows that
\begin{equation}  \label{8}
T_{\mu \nu}=S_{\mu \nu}+\frac{1}{4\kappa ^2}\left(R_0-R\right)g_{\mu \nu}.
\end{equation}
Einstein \cite{Ein2} interpreted the tensor $T_{\mu \nu}$ as \textit{the
energy-momentum tensor of "matter"}. The "matter" energy density therefore
consists of two parts, the first one originating from the electromagnetic
field, while the second one describes the gravitational contribution. From
Eq.~(\ref{8}) we obtain
\begin{equation}
S_{\mu \nu}=T_{\mu \nu}-\frac{1}{4}Tg_{\mu \nu},
\end{equation}
where $T=T_{\mu}^{\mu}$, and hence the field equation (\ref{2}) can be
reformulated as
\begin{equation}  \label{10}
R_{\mu \nu}-\frac{1}{4}g_{\mu \nu}R=\kappa ^2 \left(T_{\mu \nu}-\frac{1}{4}%
Tg_{\mu \nu}\right).
\end{equation}
In the following we will call the field equations (\ref{10}) the \textit{%
geometry-matter symmetric Einstein equations}. For a dust Universe with $%
T_0^0=\rho$, where $\rho $ is the matter density, we have $S_0^0=3\rho/4$,
and $S_i^i=-\rho/4$, $i=1,2,3$, respectively. Hence Einstein's elementary
particle theory made the \textit{remarkable prediction} that \textit{the
energy of the Universe is 75\% (3/4) electromagnetic, and 25\% (1/4)
gravitational in its origin}.

Einstein's approach to the structure of the matter, did attract very little
attention \cite{p1,p2}, even that it was sometimes briefly mentioned as a
way to solve the cosmological constant problem, by interpreting it as an
integration constant \cite{Wein}. A theory in which the electromagnetic type
energy-momentum tensor in Eq.~(\ref{1}) is substituted by the ordinary
matter energy-momentum tensor, $S_{\mu \nu}\rightarrow T_{\mu \nu}$, so that
\begin{equation}
R_{\mu \nu}+\bar{\lambda}g_{\mu \nu }R=\kappa ^2 T_{\mu \nu},
\end{equation}
was proposed by Rastall \cite{Rastall}. In this theory the matter
energy-momentum tensor is no longer conserved, and $\nabla
_{\mu}T_{\nu}^{\mu}=\lambda \nabla _{\nu}R$, $\lambda =\mathrm{constant}$.
Hence Rastall's theory implies matter creation from the gravitational field.
Different physical aspects of the Rastall theory, and its astrophysical and
cosmological implications have been investigated recently in \cite{R1,
R2,R3,R3a,R4,R5,R6,R7,R8,R9,R10}. However, in \cite{Ri} it was shown that
the Rastall theory is just a particular case of the $f(R,T)$ gravity theory
\cite{frt1}, where $T$ is the trace of the energy-momentum tensor, a modified gravity theory which is essentially based on a non-minimal geometry-matter coupling, which was
investigated in \cite{frt2,frt3,frt4,frt5,frt6,frt7,frt8,frt9,frt10,frt11,
frt12}. Several other alternative gravitational theories involving geometry-matter
couplings have also been proposed, and extensively investigated in the
physical literature, like, for example, the $f \left(R, L_m\right)$ modified
gravity theory, where $L_m$ is the matter Lagrangian \cite%
{frlm1,frlm2,frlm3, frlm4}, the Weyl-Cartan modified gravity theories \cite%
{WCW,nima}, hybrid metric-Palatini $f(R, \tilde{R})$ gravity theory \cite%
{hpm1,hpm2,hpm3}, where $\tilde{R}$ is the Ricci scalar formed from a
connection independent of the metric, the $f(R, T, R_{\mu \nu}T^{\mu \nu})$
gravity theory \cite{frtmu1,frtmu2}, or the $f(T, \tilde{T})$ gravity theory
\cite{fttt}, in which a coupling between the torsion scalar $\tilde{T}$ and
the trace of the matter energy-momentum tensor is introduced. For a review of some modified gravity models and their cosmological implications see \cite{rr9}.

The dark energy problem is a fundamental problem in present day theoretical
physics. A large number of cosmological studies, which were initiated by the
study of the high redshift Type Ia Supernovae \cite{1n,2n}, have convincingly
proven that the cosmological model according to which the Universe should
decelerate due to its own gravitational attraction is not realized in
nature. Cosmological observations have in fact convincingly shown that sometimes in the
past, at a redshift of around $z\approx 0.5$, the Universe did experience a
smooth transition to a de Sitter type accelerated expansionary phase \cite%
{1n,2n,3n,4n}. To fully explain these unexpected observational results a
deep change in our understanding of the cosmological evolution, and of its
theoretical basis, Einstein's theory of general relativity, is necessary. Hence, in
order to explain the present day observations in cosmology, and the
mysterious and puzzling large scale behavior of the Universe, many new interesting
theoretical ideas and models have been proposed recently, including loop quantum cosmology, bouncing cosmological models, string theoretical approaches, or  general scalar-tensor cosmological models with up to second-order derivatives in the field equations \cite{rr1,rr2,rr3,rr4,rr5}. Presently, from both
theoretical and observational points of view, there is an almost general
consensus in the scientific community, which we may call generally as the "standard
paradigm of the recent cosmological acceleration''  \cite{PeRa03, Pa03, rr6,rr7}. According to this new
paradigm, all cosmological observations can be easily understood and
interpreted theoretically once we assume the existence of a new and major
component in the overall content of the Universe, called \textit{dark energy} \cite{rr6,rr7}.
Hence, once the non-evolving dark energy becomes the dominant component in the Universe, it
also begins to control its evolutionary expansionary dynamics, thus determining the smooth
transition to the de Sitter type accelerated stage that characterizes the
late phases of the evolution of the Universe \cite{rr8}.  Present day observations have
also provided extremely powerful constraints on the important cosmological parameter $%
w=p/\rho $, where $p$ is the total pressure and $\rho $ is the total density
of the Universe, which provides detailed information on the temporal behavior of the
equation of state of the cosmological fluid \cite{rr8}. The analysis of the
observational data on $w$ show that it is lying somewhere in the numerical range $-1\leq w <-1/3$
\cite{acc,acc1,acc2,acc3,acc4}.  The type-Ia supernova observations, Cosmic Microwave Background, large-scale structure, Hubble measurements, and baryon acoustic oscillations show that the concordance cosmological constant model, with $w = -1$ is still safely consistent with these observational data at the 68\% confidence level \cite{acc4}. Recent results, based on full-mission Planck observations of temperature and
polarization anisotropies of the Cosmic Microwave Background Radiation
indicate, from the Planck temperature and lensing data, that the matter
density parameter of the Universe is around $\Omega _m\approx 0.31$, giving
a dark energy density parameter of the order of $\Omega _{\Lambda}\approx
0.70$ \cite{Planck}.

There are a huge number of proposals to explain the dark energy (see, for
example, the reviews \cite{PeRa03,Pa03,rr6,rr7}). One possibility is that
dark energy consists of quintessence, a slowly-varying, spatially
inhomogeneous component of the Universe \cite{8}. The idea of quintessence
can be implemented theoretically by assuming that it is the energy associated with a
scalar field $Q$, having a non-zero self-interaction potential $V(Q)$. When the
potential energy density of the scalar field is greater than the kinetic
one, the pressure $p=\dot{Q}^{2}/2-V(Q)$ associated to the scalar field $Q$%
-field is negative, thus leading to the possibility of the existence of an accelerated expansionary phase.
Several cosmological models, based on the quintessence idea, have been intensively
investigated in the physical literature (for a review see \cite{Tsu}).

The possible interaction between the dark energy and the dark matter
components of the Universe has been considered within the framework of
irreversible thermodynamics of open systems with matter
creation/annihilation in \cite{H1}. The possibility that the cosmological
anisotropy may arise as an effect of the non-comoving dark matter and dark
energy has also been investigated \cite{H2}.

Dark energy models with nonstandard scalar fields, such as phantom scalar
fields and Galileons, which can have bounce solutions and dark energy
solutions with $w<-1$ have also been considered
as an explanation of the late acceleration of the Universe \cite%
{gal1,gal2a,gal2b,gal2c, gal3,gal4,gal5}. For a review of scalar field theories
whose field equations are second order, and with second-derivative
Lagrangians see \cite{Rub}.

Scalar fields $\phi $ that are minimally coupled to gravity with a negative
kinetic energy, which are known as ``phantom fields'', have been introduced
in \cite{phan1}. The energy density and pressure of a phantom scalar field
are given by $\rho _{\phi}=-\dot{\phi}^2/2+V\left(\phi \right)$ and $p
_{\phi}=-\dot{\phi}^2/2-V\left(\phi \right)$, respectively. The properties
of phantom cosmological models have been investigated in \cite%
{phan2a,phan2b,phan2c,phan2d, phan3,phan4}.

An alternative to the scalar field dark energy models is represented by the
vector type dark energy models, and their generalizations, in which dark
energy is described by a vector field minimally coupled to gravity \cite%
{v1,v1a,v1b,v1c,v1d,v1e,v1f,v1g,v1h,v1i,v1j,v1k,v1l,v1m}, a vector field non-minimally coupled to gravity \cite{v2,v2a,v2b,v2c,v2d}, or
by some extended vector field models \cite{v3,v4}. The action for the
non-minimally massive vector field coupled to gravity is given by \cite{v2}
\begin{align}
S =-&\int d^{4}x\sqrt{-g}\Bigg[\frac{R}{2}+\frac{1}{16\pi }F_{\mu \nu
}F^{\mu \nu }-\frac{1}{2}\mu _{\Lambda }^{2}A_{\mu }A^{\mu }  \notag
\label{v1} \\
&+\omega A_{\mu }A^{\mu }R+\eta A^{\mu }A^{\nu }R_{\mu \nu }+L_{m}\Bigg],
\end{align}%
where $A^{\mu }\left( x^{\nu }\right) $, $\mu ,\nu =0,1,2,3$ is the
four-potential of the dark energy, which is allowed to couple non-minimally
to gravity. $\mu _{\Lambda }$ denotes the mass of the massive cosmological
vector field, while $\omega $ and $\eta $ are dimensionless coupling
parameters. The vector type dark energy field tensor is defined as $F_{\mu
\nu }=\nabla _{\mu }A_{\nu }-\nabla _{\nu }A_{\mu }$. So called
"superconducting" dark energy models, combining vector and scalar fields in
a gauge invariant way, were considered in \cite{S1, S2}.

It is the goal of the present paper to reformulate Einstein's theory of
\textit{"elementary particles"} as a \textit{dark energy model}. More
specifically, we assume that the vector field $S_{\mu \nu}$ in Einstein's
theory represents a vector type dark energy, and not the standard
electromagnetic field. Moreover, the model is extended from the description
of the matter particles to the entire Universe. In order to obtain the field
equation of the model, we adopt a Lagrangian that is reminiscent of the $%
f(R,T)$ theory of gravity \cite{frt1}, and contains a linear combination of the Ricci
scalar and of the trace of the energy-momentum tensor. Moreover, we assume
that the self-interacting dark energy tensor field $C_{\mu \nu}$ can be
defined in terms of the dark energy massive vector potential $\Lambda _{\mu}$%
. A coupling between the matter current and the dark energy vector potential
is also assumed.

The Einstein dark energy model field equations have the interesting property
that the matter energy-momentum tensor is not conserved. We investigate and
discuss this property from the perspective of the thermodynamics of open
systems, in which the particle number is not conserved. Therefore the
Einstein dark energy model may describe particle creation at a fundamental
level. The particle creation rates, the creation pressure, as well as the
entropy and temperature evolutions of the newly created particles are
explicitly obtained with the use of a covariant formalism.

In order to investigate the cosmological evolution of the Universe we adopt
a specific form of the dark energy vector potential. The cosmological equations, representing
generalizations of the standard Friedmann equations, are formulated in a
dimensionless form, and the redshift $z$ is adopted as the independent
variable. For the vacuum case we show that the model admits a de Sitter type
solution. We consider in detail two cosmological models, the conservative
model, in which we impose the conservation of the matter energy-momentum
tensor, and the general case, in which no specific assumptions on the
physics of the system are made. In both cases we consider in detail the
redshift evolution of the basic cosmological parameters (Hubble function,
scale factor, deceleration parameter, matter energy density, and vector
field potential, respectively), by using both analytical and numerical
techniques. In both models the Universe enters in an accelerating phase,
with the deceleration parameter taking negative values, at $z\approx 0.5$.
Depending on the numerical values of the model parameters we can obtain a
large range for the present day (negative) deceleration parameter.

The present paper is organized as follows. The variational principle of the
Einstein dark energy model is introduced in Section~\ref{sect2}, and the
corresponding field equations are derived. The thermodynamic interpretation
of the dark energy model in the framework of the irreversible thermodynamics
of open systems is discussed in Section~\ref{sect3}. The cosmological
evolution equations of the Einstein dark energy model are written down in
Section~\ref{sect4}, and their dimensionless form is also obtained. The
existence of the de Sitter solution is also proven. The general evolution of
the Universe for a specific choice of the dark energy vector potential is
investigated in Section~\ref{sect5} for both the conservative and
nonconservative cases, by using analytical and numerical techniques. It is
shown that in the presence of a self-interacting potential of the dark
energy field the Universe ends in a de Sitter phase in both conservative and
non-conservative cases. Finally, we discuss and conclude our results in
Section~\ref{sect6}.

\section{The Einstein dark energy model}\label{sect2}

In the present Section we present the basic gravitational field equations
for the Einstein dark energy model, and derive some of its basic theoretical
consequences.

\subsection{The gravitational field equations}

We assume that the Universe is filled with a cosmological vector field,
which is described by a four-potential $\Lambda _{\mu }\left( x^{\nu
}\right) $, $\mu ,\nu =0,1,2,3$. From the dark energy vector potential, in
analogy with electrodynamics, we define the dark energy field tensor
\begin{equation}
C_{\mu \nu }=\nabla _{\mu }\Lambda _{\nu }-\nabla _{\nu }\Lambda _{\mu }.
\end{equation}%
The dark energy field tensor identically satisfies the Maxwell type
equations
\begin{equation}
\nabla _{\lambda }C_{\mu \nu }+\nabla _{\mu }C_{\nu \lambda }+\nabla _{\nu
}C_{\lambda \mu }=0.  \label{13}
\end{equation}%

Furthermore, we assume that the dark energy field tensor $\Lambda_{\mu}$ is
coupled to the \textit{ordinary matter current} $j^{\mu }=\rho u^{\mu }$,
where $\rho $ is the matter energy-density, and $u^{\mu }$ is the
four-velocity of the cosmological flow, satisfying the normalization
condition $u_{\mu }u^{\mu }=-1$.

We define the energy-momentum tensor $T_{\mu \nu}$ of the matter as
\be
T_{\mu \nu}=-\frac{2}{\sqrt{-g}} \frac{\partial \left( \sqrt{-g} \mathcal{L}_{m} \right)}{%
\partial g^{\mu\nu}},
\ee
where  $\mathcal{L}_{m}$ is the Lagrangian of the total (ordinary baryonic plus dark) matter. For ordinary matter, characterized, from a thermodynamic point of view, by its energy density $%
\rho$ and thermodynamic pressure $p$ only, the energy momentum tensor is given by
\begin{equation}
T_{\mu \nu }=\left( \rho+p\right) u_{\mu}u_{\nu }+pg_{\mu \nu }.  \label{18}
\end{equation}%
In the following by $T$ we denote the trace of the matter energy-momentum tensor.

Then  the Einstein dark
energy model can be derived from the following action
\begin{align}
\tilde{S} &=\int \sqrt{-g}d^4x\Bigg[\left( 1-\beta_1 \right) \frac{R}{2\kappa^2}+%
\frac{\beta_2 }{2}T-\frac{1}{4}C_{\mu \nu }C^{\mu \nu }  \notag  \label{s1}
\\
&-\frac{m^2}{2}\Lambda _{\mu }\Lambda^\mu-V\left( \Lambda ^{2}\right) -
\frac{\alpha }{2}j^{\mu }\Lambda_\mu+\mathcal{L}_{m} \Bigg] ,
\end{align}%
where $\beta_1$, $\beta_2$, $m $ and $\alpha $ are constants.

The energy-momentum tensor $S_{\mu \nu }$ of the dark energy field,
is constructed, by analogy with classical electrodynamics, as \cite%
{LaLi}
\begin{equation}  \label{23}
S_{\mu \nu }=C_{\mu \alpha }C_{\nu }^{~\alpha }-\frac{1}{4}g_{\mu \nu
}C_{\alpha \beta }C^{\alpha \beta },
\end{equation}
with the property $S_{\mu}^{\mu}=0$.

By varying the gravitational
action with respect to the metric tensor $g^{\mu \nu}$ it follows that the cosmological evolution of the Universe
in the presence of a vector type dark energy is described by the generalized
Einstein gravitational field equations,
\bea\label{eq1}
&&\frac{(1-\beta_1)}{\kappa^2}G_{\mu\nu}-S_{\mu \nu}
-m^2\left(\Lambda_\mu\Lambda_\nu-\frac{1}{2}\Lambda^2g_{\mu\nu}\right)+\nonumber\\
&&Vg_{\mu\nu}-2V^%
\prime\Lambda_\mu\Lambda_\nu
-\frac{1}{2}\alpha \,p\,\left(u_\mu u_\nu+g_{\mu\nu}\right)\,u^\alpha\Lambda_\alpha \nonumber\\
&&=(1+\beta_2)T_{\mu\nu}-\beta_2\,\left(\mathcal{L}_m-\frac{1}{2}T\right)g_{\mu\nu},
\eea
where $V^\prime=dV/d\Lambda ^2$. In obtaining the above equation we have used the
variation of $T$ and $j^\alpha\Lambda_\alpha$ as (for the proof of these relations see \cite%
{frtmu1} and  \cite{v3}, respectively)
\begin{align}
\delta(\sqrt{-g}T)=-\sqrt{-g}\left[T_{\alpha\beta}+\left(\frac{1}{2}T-%
\mathcal{L}_m\right)g_{\alpha\beta}\right]\delta g^{\alpha\beta},
\end{align}
and
\begin{align}
\delta(&\sqrt{-g}j^\alpha\Lambda_\alpha)  \notag \\
&=\sqrt{-g}\left[j^\alpha\delta\Lambda_\alpha+\frac{1}{2}p\,\Lambda_\mu
u^\mu\,\left(u_\alpha u_\beta+g_{\alpha\beta}\right)\delta g^{\alpha\beta}\right],
\end{align}
respectively. By taking the trace of Eq.~(\ref{eq1}) we obtain
\bea
&&-\frac{\left(1-\beta _1\right)}{\kappa ^2}R+m^2\Lambda ^2+4V-2V'\Lambda ^2+\frac{3}{2}\alpha pu^{\alpha}\Lambda _{\alpha}\nonumber\\
&&=\left(3+\beta _2\right)T-4\beta _2\mathcal{L}_m .
\eea

Substituting $T$ from the
above equation into the field equations \eqref{eq1} one obtains
\begin{align}\label{17}
&R_{\mu \nu }+\,\bar\lambda Rg_{\mu \nu }=\chi T_{\mu \nu }+\bar{\kappa}%
^2S_{\mu\nu}  \notag \\
&+\bar{\kappa}^2\left\{\left(4\bar{\lambda}+1\right)Vg_{\mu\nu}+\left[2\Lambda_\mu%
\Lambda_\nu-\left(1+2\bar{\lambda}\right)\Lambda^2g_{\mu\nu}\right]V^\prime\right\}  \notag \\
&+\bar{m}^2\left(\Lambda_\mu\Lambda_\nu+\bar{\lambda}\Lambda^2g_{\mu\nu}\right)-\bar{%
\kappa}^2\frac{\left(1+2\bar{\lambda}\right)\left(1+4\bar{\lambda}\right)}{2\left(1+3\bar{\lambda}\right)}%
\mathcal{L}_mg_{\mu\nu}  \notag \\
&+\frac{1}{4}\bar{\alpha}\,p\left[2u_\mu u_\nu-\left(1+6\bar{\lambda}%
\right)g_{\mu\nu}\right]u^\alpha \Lambda_\alpha,
\end{align}
where we have defined
\begin{align}
&\quad\bar{\lambda}=-\frac{1+2\beta_2}{2(1+3\beta_2)},\quad \chi=(1+\beta_2)%
\bar{\kappa}^2,  \notag \\
&\bar{m}^2=m^2\bar{\kappa}^2,\quad\bar{\alpha}=\alpha\bar{\kappa}^2,\quad
\bar{\kappa}^2=\frac{\kappa^2}{1-\beta_1}.
\end{align}

By varying the gravitational action with respect to the vector potential of the
dark energy we obtain the equation
\begin{equation}
\bar{\kappa}^2\nabla _{\nu }C^{\nu\mu }=\frac{1}{2}\bar{\alpha}\, j^{\mu }+%
\bar{m}^2\Lambda^\mu+2\bar{\kappa}^2 V^\prime \Lambda^\mu.  \label{14}
\end{equation}
For the divergence of the energy-momentum tensor of the dark energy %
\eqref{23}, with the use of Eqs.~(\ref{13}) and (\ref{14}) we immediately
obtain,
\begin{equation}  \label{16}
\bar{\kappa}^2\nabla^{\mu}S_{\mu\nu}=\frac{1}{2}\bar{\alpha}\, C_{\nu
\mu}j^{\mu}+(\bar{m}^2+2\bar{\kappa}^2V^\prime)\,\Lambda^\alpha
C_{\nu\alpha}.
\end{equation}

\subsection{The energy and momentum conservation equations}

By taking the covariant divergence of Eq.~(\ref{17}), with the use of Eq.~(%
\ref{16}), we obtain for the divergence of the matter energy-momentum tensor
the equation
\begin{align}  \label{noncons}
\chi&\nabla^{\mu}T_{\mu \nu}=\left(\bar{\lambda}+\frac{1}{2}%
\right)\nabla_\nu R-\bar{\kappa}^2\frac{\left(1+2\bar{\lambda}\right)\left(1+4\bar{\lambda}\right)%
}{2(1+3\bar{\lambda})}\nabla_\nu\rho  \notag \\
&-\left(\bar{\lambda}+\frac{1}{2}\right)\left(\bar{m}^2-2\bar{\kappa}^2\big(%
\Lambda^2V^{\prime\prime}-V^\prime\big)\right)\nabla_\nu\Lambda^2-\frac{1}{2}%
\bar\alpha C_{\nu\alpha}j^\alpha  \notag \\
&-\frac{1}{2}\bar{\alpha}\Lambda_\nu\nabla_\mu j^\mu-\frac{1}{2}\bar{\alpha}%
\,p\,\left(u_\nu\nabla_\mu u^\mu+u_\mu\nabla^\mu u_\nu\right)\,u^\alpha\Lambda_\alpha
\notag \\
&-\frac{1}{4}\bar{\alpha}\big[2u_\mu u_\nu-\left(1+6\bar{\lambda}%
\right)g_{\mu\nu}\big]\nabla^\mu\left(u^\alpha\Lambda_\alpha p\right).
\end{align}

Let us introduce the projection operator $h_{\lambda }^{\nu}$, defined as
\begin{equation}
h_{\lambda }^{\nu}=\delta _{\lambda }^{\nu}+u_{\lambda }u^{\nu},
\end{equation}
which satisfies the condition $u_{\nu}h^{\nu}_{\lambda }= 0$. By taking the
covariant divergence of the matter energy-momentum tensor, as given by Eq.~(%
\ref{18}), we obtain first
\begin{align}  \label{25}
\nabla ^{\mu }T_{\mu \nu}=&\left(\nabla ^{\mu }\rho +\nabla ^{\mu}
p\right)u_{\mu }u_{\nu }+\left(\rho +p\right)\nabla ^{\mu} u_{\mu} u_{\nu}
\notag \\
&+\left(\rho+p\right)u_{\mu}\nabla ^{\mu }u_{\nu}+\nabla_\nu p.
\end{align}

Now, multiplying equation (\ref{noncons}) with the four-velocity $u^{\nu}$,
by taking  into account the mathematical identity $u^{\nu}\nabla ^{\mu}u_{\nu}=0$,
and using the relation \eqref{25}, we obtain the energy balance
equation in the Einstein dark energy model as
\begin{align}  \label{26}
&\left(\frac{2\chi+\bar{\alpha}\,\Theta}{2\bar{\lambda}+1}\right)\bigg[%
\dot{\rho}+3H(\rho+p)\bigg]-\bar{\kappa}^2\frac{1+4\bar{\lambda}}{1+3\bar{%
\lambda}}\,\dot{\rho}  \notag \\
&+\frac{3}{2}\bar{\alpha}\,\frac{d}{ds}(p\;\Theta)=-\dot{R}+\left[\bar{m}%
^2-2\bar{\kappa}^2(\Lambda^2V^{\prime\prime}-V^\prime)\right]\frac{d}{ds}%
\Lambda^2,
\end{align}
where we have denoted $H=(1/3)\nabla ^{\mu }u_{\mu }$, $\Theta%
=u^\alpha\Lambda_\alpha$, and the dot represents $u^{\mu }\nabla _{\mu }=d/ds$,
where $ds$ is the line element corresponding to the metric $g_{\mu \nu}$, $%
ds^2=g_{\mu \nu}dx^{\mu}dx^{\nu } $. By multiplying equation (\ref{25}) with
the projection operator $h_{\lambda }^{\nu}$ and using equation %
\eqref{noncons}, we obtain
\begin{align}  \label{27}
&u^{\mu}\nabla _{\mu }u^{\lambda }=\frac{d^2x^{\lambda }}{ds^2}+\Gamma _{\mu
\nu}^{\lambda }u^{\mu }u^{\nu }=f^\lambda,
\end{align}
where $\Gamma _{\mu \nu}^{\lambda }$ are the Christoffel symbols associated
to the metric, and the extra force $f^\lambda$ is defined as
\begin{align}
f^\lambda&=\frac{h^{\nu\lambda}}{2\chi(p+\rho)-\bar{\alpha}p\,\Theta}
\notag \\
&\Bigg\{\left(2\bar{\lambda}+1\right)\nabla_\nu R-2\chi\nabla_\nu p-\bar{\alpha}\,
C_{\nu\alpha}j^\alpha+3\bar{\alpha}H\Lambda_\nu  \notag \\
&-(1+2\bar{\lambda})\left[\bar{m}^2-2\bar{\kappa}^2\left(\Lambda^2V^{\prime%
\prime}-V^\prime\right)\right]\nabla_\nu\Lambda^2  \notag \\
&-\bar{\kappa}^2\frac{\left(1+2\bar{\lambda}\right)\left(1+4\bar{\lambda}\right)}{1+3\bar{\lambda}}%
\nabla_\nu\rho+\frac{\bar{\alpha}}{2}(1+6\bar{\lambda})\nabla_\nu\left(p\,\Theta
\right)\Bigg\}.
\end{align}
Eq.~(\ref{27}) gives the momentum balance equation for a perfect fluid in
the Einstein dark energy gravitational model. The motion of massive test particles is non-geodesic, and it takes place in the presence of a force $f^{\lambda}$ generated by the minimal coupling of the Ricci scalar with the trace of the energy-momentum tensor, and of the vector type dark energy.

\section{Thermodynamic interpretation of the Einstein dark energy model}\label{sect3}

The Einstein dark energy model has the intriguing property that the
energy-momentum tensor of the ordinary baryonic matter is \textit{not
conserved}, $\nabla^{\mu}T_{\mu \nu}\neq 0$. A non-conservative
energy-momentum tensor can naturally be interpreted in the framework of
\textit{the thermodynamics of open systems} as describing \textit{particle
creation processes}, as discussed from a theoretical fundamental point of
view in \cite{Pri0,Pri,Cal,Lima,H3,H4}. In the present Section we will
present the thermodynamic interpretation of the Einstein dark energy model
as describing irreversible particle creation in an open thermodynamic
system. In the following for the description of the Universe we adopt a
homogeneous and isotropic cosmological model, with the background geometry
given by the flat Friedmann-Robertson-Walker (FRW) metric,
\begin{equation}  \label{metr}
d s^2=-dt^2+a^2(t)\left(dx^2+dy^2+dz^2\right),
\end{equation}
where $a(t)$ is the scale factor. We adopt a comoving coordinate system with
$u^{\mu }=(1,0,0,0)$. Therefore in the FRW background geometry we have $H=%
\dot{a}/a$, a relation which defines the Hubble function, and $u^{\mu
}\nabla _{\mu }=\dot{}=d/dt$, respectively. Then Eq.~(\ref{26}) can be
rewritten in the equivalent form given by,
\begin{align}  \label{20}
\frac{d}{dt}\left(\rho a^3\right)&+p\frac{d}{dt}a^3=-\frac{a^3}{\beta}\Bigg\{%
\frac{3}{2}\bar{\alpha}\frac{d}{dt}(p\Theta)+3\bar{\kappa}^2\zeta(\rho+p)H
\notag \\
&+\dot{R}-\bigg[\bar{m}^2-2\bar{\kappa}^2\left(\Lambda^2V^{\prime\prime}-V^\prime\right)%
\bigg]\frac{d}{dt}\Lambda^2\Bigg\}.
\end{align}
where we have defined
\begin{align}
\zeta=\frac{1+4\bar{\lambda}}{1+3\bar{\lambda}},\qquad \beta=\frac{2\chi+%
\bar{\alpha}\Theta}{1+2\bar{\lambda}}-\bar{\kappa}^2\zeta.
\end{align}

\subsection{The matter creation rate and the creation pressure}

We begin our analysis by considering a thermodynamic system containing $N$
particles in a given volume $V=a^3 $. For such a system the second law of
thermodynamics in its full generality can be formulated as \cite{Pri}
\begin{equation}  \label{21}
\frac{d}{dt}\left(\rho a^3\right)+p\frac{d}{dt}a^3=\frac{dQ}{dt}+\frac{%
\mathfrak{h}}{n}\frac{d}{dt}\left(na^3\right),
\end{equation}
where $dQ$ denotes the amount of heat received by the system during time $dt$%
, $\mathfrak{h}=\rho +p$ is the enthalpy per unit volume, and $n=N/V$ is the
particle number density. Note that Eq.~(\ref{21}) represents the most
general possible formulation of the second law of thermodynamics that
explicitly takes into account not only the variation of energy, due to the
heat transfer processes, but also accounts for the energy variations due to
the change in the number of particles, an effect described by the last term $%
(\mathfrak{h}/n)d\left(na^3\right)$.

An important class of thermodynamic transformations, much used in
thermodynamics, are the adiabatic transformations, defined by the condition $%
dQ=0$. In the following we will adopt the adiabaticity condition, that is,
we will fully ignore the possible presence of proper heat transfer
mechanisms in the cosmological setting. For adiabatic transformations, in an
open system the change of the internal energy is only due to the adiabatic
variation in the number of particles. From a gravitational and cosmological
perspective, we can interpret the change of the particle number as being a
result of the energy transfer from the gravitational field to the matter.
Hence, the gravitational creation of matter may represent an important and
global source of internal energy for the Universe. In the case of adiabatic
transformations, Eq.~(\ref{21}) can be reformulated in a covariant form as
\begin{equation}  \label{30}
\dot{\rho}+3H(\rho +p)=\frac{\rho +p}{n}\left(\dot{n}+3Hn\right).
\end{equation}
A comparison of Eqs.~(\ref{26}) and (\ref{30}) shows that Eq.~(\ref{26}),
giving the energy balance equation in the Einstein dark energy model, can be
interpreted thermodynamically as describing irreversible matter creation in
an homogeneous and isotropic geometry, with the time variation of the
particle number given by the equation
\begin{equation}  \label{32}
\dot{n}+3Hn=\Gamma n.
\end{equation}
In the above equation we have defined the particle creation rate $\Gamma $
as
\begin{align}  \label{33}
\Gamma =-&\frac{1}{\beta \mathfrak{h}}\Bigg[\frac{3}{2}\,\bar{\alpha}\,\frac{%
d}{dt}(p\Theta)+3\,\bar{\kappa}^2\zeta H \mathfrak{h}  \notag \\
&+\dot{R}-\bigg(\bar{m}^2-2\bar{\kappa}^2(\Lambda^2V^{\prime\prime}-V^\prime)%
\bigg)\frac{d}{dt}\Lambda^2\Bigg].
\end{align}
Hence the energy conservation equation can be written in the form
\begin{equation}  \label{41}
\dot{\rho}+3H \mathfrak{h}=\mathfrak{h}\Gamma.
\end{equation}

In the case of adiabatic transformations, Eq.~(\ref{21}), describing
irreversible thermodynamic particle creation in open systems, can be
reformulated as an equivalent effective energy conservation equation, having
the general form
\begin{equation}
\frac{d}{dt}\left(\rho a^3\right)+\left(p+p_c\right)\frac{d}{dt}a^3=0,
\end{equation}
or, equivalently,
\begin{equation}  \label{comp}
\dot{\rho}+3\left(\rho+p+p_c\right)H=0.
\end{equation}
In the effective energy conservation laws we have introduced the new
thermodynamic function $p_c$, called \textit{the creation pressure}, and
which can be computed from the following definition \cite{Pri}
\begin{eqnarray}
p_c&&=-\frac{\rho +p}{n}\frac{d\left(na^3\right)}{da^3}=-\frac{\rho +p}{3nH}%
\left(\dot{n}+3nH\right)  \notag \\
&&=-\frac{\rho+p}{3}\frac{\Gamma }{H}=-\frac{1}{3}\frac{\mathfrak{h}}{H}%
\Gamma.
\end{eqnarray}
Note that the creation pressure is proportional to the particle creation
rate $\Gamma$. In the Einstein dark energy model, the creation pressure is
determined by the equation
\begin{align}  \label{pc}
p_c=&\frac{1}{3\beta H}\Bigg[\frac{3}{2}\,\bar{\alpha}\,\frac{d}{dt}\left(p\Theta
\right)+3\,\bar{\kappa}^2\zeta H \mathfrak{h}  \notag \\
&+\dot{R}-\bigg(\bar{m}^2-2\bar{\kappa}^2(\Lambda^2V^{\prime\prime}-V^\prime)%
\bigg)\frac{d}{dt}\Lambda^2\Bigg].
\end{align}

\subsection{Entropy production and temperature evolution}

For an arbitrary physical system, its entropy $S$, whose variation is
governed by the second law of thermodynamics, can be usually decomposed into
two terms: the first one is the entropy change due to the presence of an
entropy flow $d_eS$, while the second one corresponds to the entropy $d_iS$
created within the system. Hence the variation of the total entropy $S$ of a
physical system is given by the fundamental relation \cite{Pri0,Pri}
\begin{equation}
dS = d_eS + d_iS,
\end{equation}
with $d_iS$ always satisfying the condition $d_iS \geq 0$, as imposed by the
second law of thermodynamics. On the other hand the entropy flow as well as
the entropy production term can be estimated from the total differential $dS$
of the entropy, which is given by \cite{Pri},
\begin{equation}  \label{211}
\mathcal{T} d\left(sa^3\right)=d\left(\rho a^3\right)+pda^3-\mu d\left(na^3\right),
\end{equation}
where $\mathcal{T}$ is the temperature of the system, $s=S/a^3$ is the entropy per
unit volume, and $\mu $ is the chemical potential, defined by the relation
\begin{equation}  \label{chemi}
\mu n=\mathfrak{h}-\mathcal{T}s.
\end{equation}

For adiabatic transformations and for closed systems, the second law of
thermodynamics imposes the conditions $dS=0$ and $d_iS=0$. However, in the
presence of matter creation taking place in open thermodynamic systems, even
in the case of adiabatic transformations with no heat exchange there still
is a non-zero contribution to the entropy. Note that for homogeneous
thermodynamic open systems the condition $d_eS = 0$ also must hold. On the
other hand, for open systems, matter creation gives a significant
contribution to the entropy production, with the time variation of the
entropy satisfying the relation \cite{Pri}
\begin{align}  \label{43}
\mathcal{T}\frac{d_iS}{dt}&=\mathcal{T}\frac{dS}{dt}=\frac{\mathfrak{h}}{n}\frac{d}{dt}%
\left(na^3\right)-\mu \frac{d}{dt}\left(na^3\right)  \notag \\
&=\mathcal{T}\frac{s}{n}\frac{d}{dt}\left(na^3\right)\geq 0,
\end{align}
where we have used the relations given by \eqref{21} and \eqref{211}. From
Eq.~(\ref{43}) we obtain the entropy variation in an open system due to
particle production as
\begin{equation}  \label{43a}
\frac{dS}{dt}=\frac{S}{n}\left(\dot{n}+3Hn\right)=\Gamma S\geq 0.
\end{equation}
With the use of Eq.~(\ref{32}), giving the particle number balance in the
Einstein dark energy model, we obtain for the entropy production rate of the
Universe, the equation
\begin{align}  \label{entfL}
\frac{1}{S}\frac{dS}{dt}=&-\frac{1}{\beta \mathfrak{h}}\Bigg[\frac{3}{2}\,%
\bar{\alpha}\,\frac{d}{dt}(p\Theta)+3\,\bar{\kappa}^2\zeta H \mathfrak{h}
\notag \\
&+\dot{R}-\bigg(\bar{m}^2-2\bar{\kappa}^2(\Lambda^2V^{\prime\prime}-V^\prime)%
\bigg)\frac{d}{dt}\Lambda^2\Bigg].
\end{align}

Another important thermodynamic quantity, the entropy flux vector $S^{\mu }$%
, is defined according to \cite{Cal}
\begin{equation}
S^{\mu }=n\sigma u^{\mu },
\end{equation}
where $\sigma =S/N$ denotes the entropy per particle, and $u^\mu$ is the
particle four-velocity. The entropy flux four-vector must satisfy the second
law of thermodynamics, which imposes the condition $\nabla _{\mu }S^{\mu
}\geq 0$. On the other hand the entropy per particle $\sigma $  must also satisfy the
Gibbs equation \cite{Cal},
\begin{equation}  \label{47}
n\mathcal{T}d\sigma =d\rho-\frac{\mathfrak{h}}{n}dn.
\end{equation}
The chemical potential $\mu $ can be obtained from Eq.~\eqref{chemi} as
\begin{equation}\label{mu}
\mu =\frac{\mathfrak{h}}{n}-\mathcal{T}\sigma .
\end{equation}

By using the definition of the chemical potential $\mu$ as given by Eq.~(\ref{mu}), we immediately obtain
\begin{align}  \label{49}
\nabla _{\mu }S^{\mu }&=\left( \dot{n}+3nH\right) \sigma +n\dot{\sigma}
\notag \\
&=\frac{1}{\mathcal{T}}\left( \dot{n}+3Hn\right) \left( \frac{\rho +p}{n}-\mu \right)=%
\frac{\Gamma }{\mathcal{T}}\left(\mathfrak{h}-\mu n\right) .  \notag \\
\end{align}
In obtaining Eq.~(\ref{49}) we have also taken into account the relation
\begin{eqnarray}
n\mathcal{T}\dot{\sigma}=\dot{\rho}-\frac{\rho+p}{n}\dot{n}=0,
\end{eqnarray}
which follows immediately from Eqs. \eqref{43a} and \eqref{47}. With the use
of Eqs.~(\ref{32}) and (\ref{33}) we obtain for the entropy production rate,
associated to the particle production processes in the Einstein dark energy
model the expression
\begin{align}
\nabla _{\mu }S^{\mu }&=-\frac{1}{\beta \mathcal{T}} \left(1-\frac{\mu n}{\rho +p}
\right)\Bigg[\frac{3}{2}\,\bar{\alpha}\,\frac{d}{dt}(p\Theta)+3\,\bar{%
\kappa}^2\zeta H \mathfrak{h}  \notag \\
&+\dot{R}-\bigg(\bar{m}^2-2\bar{\kappa}^2(\Lambda^2V^{\prime\prime}-V^\prime)%
\bigg)\frac{d}{dt}\Lambda^2\Bigg].
\end{align}
Finally, we consider the general case of a perfect comoving cosmological
fluid that can be described by only two essential thermodynamic parameters,
namely, the particle number density $n$, and the temperature $\mathcal{T}$. In this
case the energy density $\rho$ and the thermodynamic pressure $p$ can be
obtained in terms of $n$ and $\mathcal{T}$ with the use of the equilibrium equations
of state,
\begin{equation}  \label{51}
\rho =\rho \left(n,\mathcal{T}\right),\qquad p=p\left(n,\mathcal{T}\right).
\end{equation}
Hence the energy conservation equation (\ref{41}) takes the form
\begin{equation}
\left(\frac{\partial \rho }{\partial n}\right)_{\mathcal{T}}\dot{n}+\left(\frac{\partial
\rho }{\partial \mathcal{T}}\right)_n\dot{\mathcal{T}}+3H\left(\rho +p\right)=\mathfrak{h}\Gamma.
\end{equation}
With the use of the general thermodynamic relation \cite{wein}
\begin{equation}
\left(\frac{\partial \rho }{\partial n}\right)_{\mathcal{T}}=\frac{\rho +p}{n}-\frac{\mathcal{T}}{n%
}\left(\frac{\partial p}{\partial \mathcal{T}}\right)_n,
\end{equation}
we obtain the temperature evolution of the newly created particles as a
function of the particle creation rate and the speed of sound $%
c_s^2=(\partial p/\partial \rho)_n $ as
\begin{align}  \label{55}
\frac{\dot{\mathcal{T}}}{\mathcal{T}}=\frac{\dot{n}}{n}\left(\frac{\partial p}{\partial \rho }%
\right)_n=\left(\Gamma -3H\right)c_s^2.
\end{align}

Hence in the case of the Einstein dark energy model for the temperature
variation of the cosmological matter we obtain the relation
\begin{align}
\frac{\dot{\mathcal{T}}}{\mathcal{T}}=-\frac{c_s^2}{\beta \mathfrak{h}}\Bigg\{&\frac{3}{2}\bar{%
\alpha}\frac{d}{dt}(p\Theta)+3\bar{\kappa}^2\zeta H\mathfrak{h}+\dot{R}%
+3\beta H\mathfrak{h}  \notag \\
&-\bigg[\bar{m}^2-2\bar{\kappa}^2\left(\Lambda^2V^{\prime\prime}-V^\prime%
\right)\bigg]\frac{d}{dt}\Lambda^2\Bigg\}.
\end{align}

\section{Cosmological dynamics in the Einstein dark energy model}

\label{sect4}

In the present Section we investigate the cosmological implications of the
Einstein dark energy model. We assume that the Universe is isotropic and
homogeneous, and that its large scale geometry is described by the flat FRW
metric (\ref{metr}).

In order to facilitate the comparison with the observational data we also
introduce the deceleration parameter $q$, which is an essential indicator of
the accelerating/decelerating expansion, and is defined as
\begin{equation}
q=\frac{d}{dt}\frac{1}{H}-1.  \label{deccparam}
\end{equation}
Negative values of $q$ correspond to accelerating evolution, whereas
positive values indicate deceleration. In order to obtain cosmological
results that can be compared easily with the astronomical observations
instead of the time variable $t$ we introduce as independent variable the
redshift $z$, defined as
\begin{equation}
1+z=\frac{1}{a},
\end{equation}%
where we have normalized the scale factor so that its present day value is
one, $a(0)=1$. Therefore we obtain
\begin{equation}  \label{61a}
\frac{d}{dt}=\frac{dz}{dt}\frac{d}{dz}=-(1+z)H(z)\frac{d}{dz}.
\end{equation}%
As a function of the cosmological redshift the deceleration parameter $q$
can be obtained as
\begin{equation}
q(z)=(1+z)\frac{1}{H(z)}\frac{dH(z)}{dz}-1.
\end{equation}

\subsection{Cosmology of the geometry-matter symmetric Einstein model}

The geometry-matter symmetric Einstein theory is described by Eqs.~(\ref{10}%
). For a flat geometry described by the FRW metric (\ref{metr}), we have $R_0^0=3\left(\dot{H}+H^2\right)$, $R_i^i=\dot{H}+3H^2$, $i=1,2,3$, and $R=6\left(\dot{H}+2H^2\right)$, respectively. Hence the full
set of the cosmological field equations (\ref{10}) reduces to a {\it single} cosmological evolution equation, given by
\begin{equation}  \label{60}
\dot{H}=-\frac{\kappa ^2}{2}\left(\rho+p\right).
\end{equation}
By taking the covariant divergence of Eq.~(\ref{10}) gives
\be
\frac{1}{4}\nabla _{\nu}R=\kappa^2\left(\nabla _{\mu}T^{\mu}_{\nu}-\frac{1}{4}\nabla _{\nu}T\right).
\ee
In the case of the Friedmann-Robertson-Walker metric (\ref{metr}), for a Universe filled with an ideal cosmological fluid we obtain the second cosmological evolution equation of the geometry-matter symmetric theory as
\begin{equation}  \label{61}
\dot{\rho}+3\left(\rho+p\right)H=-\frac{1}{4}\left(\frac{\dot{R}}{\kappa ^2}+%
\dot{T}\right),
\end{equation}
respectively. With the use of Eq.~(\ref{60}), Eq.~(\ref{61}) can be
immediately integrated to give
\begin{equation}
\rho=\frac{3}{\kappa ^2}H^2-\frac{1}{4}\left(\frac{R}{\kappa ^2}%
+T\right)+\rho _0,
\end{equation}
where $\rho _0$ is an arbitrary integration constant. When $R/\kappa ^2 +T=%
\mathrm{constant}$, we recover standard general relativity in the presence
of an arbitrary integration constant, which can be interpreted as the
cosmological constant. The system of Eqs.~(\ref{60}) and (\ref{61}) has as a
unique vacuum solution with $\rho =p=0$ the de Sitter type exponential
solution, with $H=H_0=\mathrm{constant}$, $a(t)=a_0\exp\left (H_0t\right)$,
and $R=\mathrm{constant}=12H_0^2$, respectively.

For the case of the power-law expansion of the Universe, with $%
a=a_0\left(t/t_0\right)^{\varpi }$, $\varpi =\mathrm{constant}$, we have $%
H=\varpi /t$. By assuming the equation of state of the form $p=\left(\gamma
-1\right)\rho$, with $\rho =\rho _0/t^2$, $\gamma \in [1,2]$, $\rho _0=%
\mathrm{constant}$, Eq.~(\ref{60}) is identically satisfied for
\begin{equation}
\varpi =\frac{\kappa ^2 \gamma \rho _0}{2}.
\end{equation}
With the choice $\rho _0=4/(3\gamma ^2\kappa ^2)$, we recover the solutions
of the standard general relativity with $a(t)=a_0(t/t_0)^{2/3\gamma}$.
However, in the geometry-matter symmetric Einstein cosmological model more
general power law solutions are also possible.

Finally, we consider that the scale factor is given by $a(t)=a_0\sinh
\left(H_0t\right)$, where $H_0$ is a constant. For the linear barotropic
equation of state $p=\left(\gamma -1\right)\rho$, Eq.~(\ref{60}) gives
\begin{equation}
\rho (t)=\frac{2H_0^2}{\gamma \kappa ^2}\frac{1}{\sinh ^2\left(H_0t\right)}.
\end{equation}
The Ricci scalar of this model is given by $R(t)=6 H_0^2 \cosh \left(2 H_0
t\right) \text{csch}^2\left(H_0 t\right)$. The energy density is singular at
$t=0$, and in the large time limit it ends in a vacuum state, after a
transition to an accelerating, de Sitter type phase.

\subsection{Cosmological evolution equations of the vector Einstein dark
energy model}

In the following we will investigate the cosmology of the Einstein dark
energy model in the presence of a vector field. From the homogeneity and
isotropy of the Universe it follows that the dark energy vector field $%
\Lambda _{\mu}$ can have only a time component, which should be a function of
the cosmic time $t$ only, so that
\begin{equation}
\Lambda _{\mu}=\Lambda_{\mu}(t)=\left(-\Lambda_0(t),0,0,0\right).
\end{equation}
The (00) and (ii) components of the Einstein field equation \eqref{17} gives the cosmological evolution  equations as
\begin{align}  \label{63}
&12\,(1+4\bar{\lambda})H^2+12(1+2\bar{\lambda})\dot{H}-3\bar{\alpha}(1+2\bar{%
\lambda})\Lambda_0\,p  \notag \\
&+4\big[(1+\bar{\lambda})\bar{m}^2+\bar{\kappa}^2(1-2\bar{\lambda})V^\prime%
\big]\Lambda_0^2-4\bar{\kappa}^2(1+4\bar{\lambda})V  \notag \\
&+2\left[2\chi-\bar{\kappa}^2\zeta(1+2\bar{\lambda})\right]\,\rho=0,
\end{align}%
\begin{align}  \label{64}
&12\,(1+4\bar{\lambda})H^2+4(1+6\bar{\lambda})\dot{H}-\bar{\alpha}(1+6\bar{%
\lambda})\Lambda_0\,p  \notag \\
&+4\big[\bar{\lambda}\bar{m}^2-\bar{\kappa}^2(1+2\bar{\lambda})V^\prime\big]%
\Lambda_0^2-4\bar{\kappa}^2(1+4\bar{\lambda})V  \notag \\
&-2\bar{\kappa}^2\zeta(1+2\bar{\lambda})\,\rho-4\chi \,p=0,
\end{align}
and the vector field equation of motion \eqref{14} will be reduced to
\begin{align}  \label{65}
\big(\bar{m}^2+2\bar{\kappa}^2V^\prime\big)\Lambda_0=-\frac{1}{2}\bar{\alpha}%
\,\rho.
\end{align}
The temporal component of the conservation equation \eqref{16} reads
\begin{align}  \label{66}
&\left[2\bar{\kappa}^2\zeta(1+2\bar{\lambda})-4\chi\right]\dot{\rho}-2\bar{%
\alpha}\Lambda_0\dot{\rho}  \notag \\
&+3\bar{\alpha}(1+2\bar{\lambda})(\dot{p}\Lambda_0+p\dot{\Lambda}%
_0)-12\chi(\rho+p)H  \notag \\
&-4(1+2\bar{\lambda})\Lambda_0\dot{\Lambda}_0\big[2\bar{\kappa}%
^2V^{\prime\prime}\Lambda_0^2+\bar{m}^2+2\bar{\kappa}^2V^\prime\big]  \notag
\\
&-6\bar{\alpha}\Lambda_0(\rho-p)H-12(1+2\bar{\lambda})(4H\dot{H}+\ddot{H})=0.
\end{align}

After substituting $V^{\prime }$ from Eq.~(\ref{65}) into Eqs.~(\ref{63})
and (\ref{64}), respectively, we can solve them algebraically to obtain $%
H^{2}$ and $\dot{H}$ as
\begin{eqnarray}  \label{67}
H^{2}&=&\frac{\left[ -\bar{\alpha}\left( 1+3\bar{\lambda}\right) \Lambda
_{0}+\bar{\kappa}^{2}\left( 1+2\bar{\lambda}\right) \right] \rho }{6\left(
1+3\bar{\lambda}\right) }+  \notag \\
&&\frac{\chi \left[ \left( 1+6\bar{\lambda}\right) \rho +3\left( 1+2\bar{%
\lambda}\right) p\right] }{6\left( 1+4\bar{\lambda}\right) }+\frac{2\bar{%
\kappa}^{2}V-\Lambda _{0}^{2}\bar{m}^{2}}{6},  \notag \\
\end{eqnarray}%
and
\begin{equation}  \label{68}
\dot{H}=-\frac{1}{4}(2\chi-\bar{\alpha }\Lambda _{0})\left(\rho+ p \right) ,
\end{equation}%
respectively. Eqs.~(\ref{67}) and (\ref{68}) represents the generalization of the standard Friedmann equations for the Einstein dark energy model. In the standard general relativistic limit with $\beta _1=\beta _2=\alpha =m=V=0$, and in the absence of the vector field $C^{\mu \nu}$, we have $\bar{\lambda}=-1/2$, and Eqs.~(\ref{67}) and (\ref{68}) reduce to the general relativistic Friedmann equations $3H^2=\rho$, and $\dot{H}=-\left(\kappa ^2/2\right)\left(\rho +p\right)$, respectively.

\subsection{The de Sitter solution}

As a first example of the cosmological evolution in the Einstein dark energy
model we consider the case of the vacuum Universe, with $\rho =p=0$. In this
case Eq.~(\ref{65}) can be immediately integrated to give for the dark
energy self-interaction potential the expression
\begin{equation}  \label{69a}
V\left(\Lambda _0^2\right)=V_0+\frac{\bar{m}^2}{2\bar{\kappa}^2}\Lambda _0^2,
\end{equation}
where $V_0$ is an arbitrary integration constant. The first term in the
above equation is the cosmological constant, while the second term cancels
the mass term in the action. Hence, the cosmological evolution of the
Universe is independent on the time component $\Lambda _0$ of the vector
dark energy field.

Moreover, we assume that in this vacuum phase the Hubble function is
becoming a constant, so that $H=H_0=\mathrm{constant}$. Then the energy
conservation Eq.~(\ref{66}) is identically satisfied, while Eqs.~(\ref{63})
and (\ref{64}) reduce to a single algebraic equation, given by
\begin{equation}
\left(1+4 \bar{\lambda} \right) \left(3 H_0^2-\bar{\kappa} ^2 V_0\right)=0,
\end{equation}
which immediately gives
\begin{equation}
H_0=\sqrt{\frac{\bar{\kappa} ^2V_0}{3}}.
\end{equation}

Therefore the Einstein dark energy model does admits a de Sitter type vacuum
exponentially expanding solution.



\subsection{Dimensionless form of the cosmological evolution equations}

In order to simplify the mathematical formalism we introduce a set of
dimensionless variables $\left( \tau ,h,r,P,v,\lambda_0,\xi\right) $,
defined as
\begin{align}  \label{69}
&\tau =H_{0}t,\quad H=H_{0}h,\quad\rho =\frac{3H_{0}^{2}}{\bar{\kappa} ^{2}}%
r,\quad p=\frac{ 3H_{0}^{2}}{\bar{\kappa} ^{2}}P,  \notag \\
&\qquad V =\frac{H_{0}^{2}}{2\bar{\kappa} ^{2}}v,\quad
\Lambda_0=H_0\lambda_0,\quad \bar{\alpha}=\bar{\kappa}^2H_0^{-1}\xi,
\end{align}%
where $H_{0}$ is the present day value of the Hubble function. Hence the
cosmological field equations of the Einstein dark energy model can be
formulated in a dimensionless form as
\begin{eqnarray}  \label{67a}
h^{2} &=&\frac{\left[ -\xi \left( 1+3\bar{\lambda}\right) \lambda
_{0}+\left( 1+2\bar{\lambda}\right) \right] r}{2\left( 1+3\bar{\lambda}%
\right) }+  \notag \\
&&\frac{\left[ \left( 1+6\bar{\lambda}\right) r+3\left( 1+2\bar{\lambda}%
\right) P\right] }{4\left( 1+3\bar{\lambda}\right) }+\frac{v-\lambda _{0}^{2}%
\bar{m}^{2}}{6},  \notag \\
\end{eqnarray}%
\begin{equation}  \label{68a}
\frac{dh}{d\tau }=-\frac{3}{4}\left[ \zeta -\xi \lambda _{0}\right] \left(
r+P\right) ,
\end{equation}%
\begin{align}  \label{66a}
& 4\bar{\lambda}\zeta \frac{dr}{d\tau }-2\xi \lambda _{0}\frac{dr}{d\tau }%
+3\xi (1+2\bar{\lambda})\frac{d}{d\tau }\left( P\lambda _{0}\right)  \notag
\\
& -6\zeta (r+P)h-\frac{4}{3}(1+2\bar{\lambda})\lambda _{0}\frac{d\lambda _{0}%
}{d\tau }\big(\lambda _{0}^{2}v^{\prime \prime }+v^{\prime }+\bar{m}^{2}\big)
\notag \\
&- 6\xi \lambda _{0}(r-P)h-4(1+2\bar{\lambda})\left(4h\frac{dh}{d\tau }+%
\frac{d^{2}h}{d\tau ^{2}}\right)=0,
\end{align}
\begin{equation}  \label{rd}
\frac{3}{2}\xi r=-\lambda _{0}\left( v^{\prime }+\bar{m}^{2}\right),
\end{equation}
and
\begin{equation}  \label{eos}
P=P(r),
\end{equation}
where we have defined $v^{\prime}=dv/d(\lambda_\mu\lambda^\mu)$ and $%
v^{\prime\prime}=d^2v/d(\lambda_\mu\lambda^\mu)^2$. The system of equations (\ref{67a})-(\ref{eos}) must be
integrated with the initial conditions $h\left(\tau_{pres}\right)=1$, $%
r\left(\tau_{pres}\right)=1$, and $\lambda
_0\left(\tau_{pres}\right)=\lambda _0^{(0)}$, respectively, where $%
\tau_{pres}$ is the present age of the Universe.

For the sake of comparison we also consider the behavior of the cosmological parameters in the standard $\Lambda $CDM model. By assuming that the Universe is filled with dust matter only, with negligible pressure, the energy conservation equation $\dot{\rho}+3H\rho=0$ gives for the variation of matter density the expression $\rho\sim 1/a^3\sim (1+z)^3$. The evolution of the Hubble function is given by \cite{Planck}
\be
H=H_0\sqrt{\left(\Omega _{DM}+\Omega _b\right)a^{-3}+\Omega _{\Lambda}},
\ee
where $\Omega _{DM}$, $\Omega _b$ and $\Omega _{\Lambda}$ are the density parameters of the cold dark matter, baryonic matter, and dark energy (a cosmological constant), respectively, satisfying the constraint $\Omega _{DM}+\Omega _b+\Omega _{\Lambda}=1$. As a function of the redshift the Hubble function can be written in a dimensionless form as
\be
h(z)=\sqrt{\left(\Omega _{DM}+\Omega _b\right)\left(1+z\right)^{3}+\Omega _{\Lambda}}.
\ee

For the redshift dependence of the deceleration parameter we find
\be
q(z)=\frac{3 (1+z)^3 \left(\Omega _{DM}+\Omega _b\right)}{2 \left[\Omega _{\Lambda}+(1+z)^3
   \left(\Omega _{DM}+\Omega _b\right)\right]}-1.
\ee
In the following for the density parameters we adopt the values $\Omega _{DM}=0.2589$, $\Omega _{b}=0.0486$, and $\Omega _{\Lambda}=0.6911$ \cite{Planck}, giving for the total matter density parameter the value $\Omega _m=\Omega _{DM}+ \Omega _b$ the value $\Omega _m=0.3089$. With the use of these numerical values of the cosmological parameters we obtain the present day value of the deceleration parameter as $q(0)=-0.5381$. On the other hand for the variation of the dimensionless matter density with respect to the redshift we obtain the expression $r(z)=\Omega _m(1+z)^3=0.3089(1+z)^3$.

\section{General cosmological evolution of the Universe }
\label{sect5}

In the present Section we consider the general cosmological dynamics of the
Universe in the Einstein dark energy model. We will investigate two types of
model, corresponding to the conservative evolution of the Universe, in which
the energy density of the matter satisfies the usual conservation equation,
as well as the general case in which the matter energy-momentum tensor is
not conserved.

\subsection{The cosmological evolution of the conservative Einstein dark
energy model}

In the following we analyze the case of the conservative Einstein dark
energy models. Hence we assume that the total matter energy density in the
Universe is conserved, and therefore $\rho $ satisfies the conservation
equation
\begin{equation}
\dot{\rho}+3H(\rho +p)=0.
\end{equation}%
With the use of the condition of the matter conservation, and by
substituting $\ddot{H}$ as obtained from Eq.~(\ref{68}), Eq.~(\ref{66})
gives for the time evolution of the time component of the Einstein vector
field the equation
\begin{align}
\frac{d\Lambda _{0}}{dt}&=-\frac{3H}{\left( 1+2\bar{\lambda}\right) \left(
1+3\bar{\lambda}\right) \left( 8\bar{\kappa}^{2}\Lambda _{0}^{3}V^{\prime
\prime }+\bar{\alpha}\rho \right) }\times  \notag \\
&\Bigg\{ \left( 1+2\bar{\lambda}\right) \rho \Big[2\bar{\kappa}^{2}\left( 1+4%
\bar{\lambda}\right) +\left( 1+3\bar{\lambda}\right)\bar{\alpha} \Lambda
_{0} \Big]  \notag \\
& + p\Big[2\bar{\kappa}^2(1+2\bar{\lambda})(1+4\bar{\lambda})-(3-2\bar{%
\lambda})(1+3\bar{\lambda})\bar{\alpha}\Lambda_0\Big]\Bigg\}  \notag \\
&+\frac{6\chi \left[ H(p+\rho )+\dot{p}\right] }{8\bar{\kappa}^{2}\Lambda
_{0}^{3}V^{\prime \prime }+\bar{\alpha}\rho }.
\end{align}

Since the constant $\beta _1$ just changes the magnitude of the
gravitational coupling constant, which is tightly constrained by experiment,
we will take it as zero, $\beta _1=0$, so that
\begin{equation}
\bar{\kappa }^2=\kappa ^2.
\end{equation}

Then in the dimensionless variables introduced in Eq.~(\ref{69}) the basic
equations describing the cosmological evolution of the conservative Einstein
dark energy model take the form
\begin{equation}  \label{72}
\frac{dh}{d\tau }=-\frac{3}{4}\left( \zeta-\xi\lambda _{0}\right) \left(
r+P\right) ,
\end{equation}%
\begin{align}  \label{evol}
&\frac{d\lambda _{0}}{d\tau}=\frac{9\zeta \left[ h(P+r )+\frac{dP}{d\tau}%
\right] }{4\lambda _{0}^{3}v^{\prime \prime }+3\xi r}  \notag \\
&-\frac{9h}{ 4\lambda _{0}^{3}v^{\prime \prime }+3\xi r } \Bigg( \Big[%
2\zeta+\xi\lambda _{0} \Big]r+ \Big[2\zeta-\frac{14-11\zeta}{2-\zeta}%
\xi\lambda_0\Big]P\Bigg),
\end{align}
\begin{equation}  \label{11}
\left( \bar{m}^{2}+v^{\prime }\right)\lambda _{0}=-\frac{3}{2}\xi\, r ,
\end{equation}%
\begin{equation}  \label{cons}
\frac{dr}{d\tau }+3h\left( r+P\right) =0.
\end{equation}

In the following we will assume that the matter content of the Universe
satisfies the linear barotropic equation of state
\begin{equation}  \label{12}
p=\left( \gamma -1\right) \rho ,
\end{equation}%
where $1\leq \gamma \leq 2$ is a constant. \newline

\subsubsection{The self-interaction potential of the vector field in the
conservative Einstein dark energy model}

By differentiating equation \eqref{11} and using the conservation equation %
\eqref{cons}, one can obtain
\begin{align}  \label{15}
\frac{d\lambda _{0}}{d\tau}=\frac{9\,\xi\, h\, (r+P)}{2(\bar{m}%
^2+v^\prime-2\lambda_0^2v^{ \prime\prime})}.
\end{align}
Now, equating equations \eqref{evol} and \eqref{15}, one can obtain a
differential equation for $v(\lambda_0^2)$ as
\begin{align}  \label{82}
&\Bigg(\frac{\gamma(3\gamma-2)(\zeta-2)\zeta-\lambda_0\big[%
16-14\gamma+(11\gamma-12)\zeta\big]\xi}{\lambda_0(\zeta-2)\,\xi}  \notag \\
&\qquad+\gamma \Bigg)\left(\bar{m}^2+v^\prime\right)=0,
\end{align}
where we have used equations \eqref{11} and \eqref{12} to eliminate $P$ and $%
r$ in favor of $\lambda_0$. There are two possibilities for the above
equation to be satisfied, corresponding to the vanishing of each parenthesis
independently, as well as to their simultaneous cancellation. The second
parenthesis gives
\begin{align}
v=v_0+\bar{m}^2\lambda_0^2.
\end{align}
The first term is the cosmological constant, which will produce the
accelerating expansion for the universe. The second term will cancel the
mass term in the action. Once the condition $\bar{m}^2+v^\prime=0$ is
satisfied, we immediately obtain $r=P=0$, and Eq.~(\ref{15}) gives $\lambda
_0=\mathrm{constant}$. Then from Eq.~(\ref{72}) have $h=\mathrm{constant}$,
and therefore we have reobtained the de Sitter type solution of the Einstein
dark energy model. The cancellation of the first parentheses gives for $%
\lambda _0$ the expression
\begin{equation}
\lambda _0=\mathrm{constant},
\end{equation}
leading again, via Eqs.~(\ref{15}) and (\ref{72}) to $r=P=0$ and $h=\mathrm{%
constant}$, respectively.

Therefore, as a conclusion of this analysis we can
state that the Einstein dark energy model does not have conservative non-vacuum
solutions.

\subsection{The non-conservative cosmological evolution in the Einstein dark
energy model}

Next we consider the evolution of the Universe in the non-conservative
version of the Einstein dark energy model. We assume again that the matter
content of the Universe satisfies the linear barotropic equation of state (%
\ref{12}). The matter energy density is obtained from \eqref{rd} as
\begin{equation}
r\left(\lambda _0\right)=-\frac{2}{3\xi}\lambda _0\left(v^\prime+\bar{m}%
^2\right),
\end{equation}
giving
\begin{equation}  \label{110}
\frac{dr}{d\tau}= -\frac{2}{3\xi}\left(-2\lambda _0^2v^{\prime \prime
}+v^{\prime }+\bar{m}^2\right)\frac{d\lambda _0}{d\tau}.
\end{equation}

With the use of the linear barotropic equation of state Eq.~(\ref{68a})
becomes
\begin{equation}  \label{111}
\frac{dh}{d\tau }=\frac{\gamma}{2\xi}\left[\zeta -\xi \lambda _{0}\right]
\lambda _0\left(v^\prime+\bar{m}^2\right),
\end{equation}
while Eq.~(\ref{67a}) takes the form
\begin{equation}
h^2=\left[-\frac{1}{2}\xi \lambda _0+\zeta -1-\frac{3\gamma }{4}\left(\zeta
-2\right)\right]r+\frac{v-\lambda _0^2\bar{m}^2}{6}.
\end{equation}
By taking the derivative with respect to $\tau $ of the above equation, and
substituting $dr/d\tau$ and $dh/d\tau $ as given by Eqs.~(\ref{110}) and (%
\ref{111}), respectively, we obtain the time evolution equation of $\lambda
_0$ as
\begin{align}\label{113}
\frac{d\lambda _0}{d\tau}=\frac{-3\gamma \left(\zeta -\xi \lambda _0\right)\lambda _0\left(v'+\bar{m}^2\right)h}{2\left[\zeta-\f{\xi \lambda _0}{2} -1-\f{3\gamma}{4}(\zeta -2)\right]\left(\bar{m}^2+v'-2\lambda _0^2v''\right)}.
\end{align}

The system of ordinary coupled differential equations (\ref{111}) and (\ref%
{113}) gives the general cosmological evolution of the Universe in the
Einstein dark energy model. In order to simplify the field equations we
rescale the potential $\lambda _0$ by introducing a new variable $\theta $
defined as $\theta =\xi \lambda _0$.

In order to close the system of field equations we need to fix the
functional form of the self-interaction potential of the vector dark energy
field, which we assume to be
\begin{equation}
v\left( \xi ^{2}\lambda _{0}^{2}\right) =v_{0}+\nu \left( \xi ^{2}\lambda _0
^{2}\right) ^{2},
\end{equation}%
%
%
%
%
%
%
where $v_{0}\geq 0$ and $\nu \geq 0$ are constants. Then we immediately
obtain $v^{\prime }=-2\nu \xi ^{2}\left( \xi ^{2}\lambda _{0}^{2}\right) $,
and $v^{\prime \prime }\left( \lambda _{0}^{2}\right) =2\nu \xi ^{4}$,
respectively.

Hence Eqs.~(\ref{111}) and (\ref{113}) can be written as
\begin{equation}  \label{111a}
(1+z)h\frac{dh}{dz}=\nu \gamma \theta \left( \zeta-\theta \right) \left(
\theta ^{2}-\tilde{k}^{2}\right) ,
\end{equation}%
and
\begin{equation}  \label{116}
(1+z)\frac{d\theta }{dz}=\frac{3\gamma \theta \left( \zeta -\theta \right)
\left( \theta ^{2}-\tilde{k}^{2}\right) }{-3\theta ^{3}+3\omega \theta ^{2}+%
\tilde{k}^{2}\theta -\omega \tilde{k}^{2}},
\end{equation}%
where we have denoted
\begin{equation}
\tilde{k}^{2}=\frac{\bar{m}^{2}}{2\nu \xi ^{2}},\omega =2\left[ \zeta -1-%
\frac{3\gamma }{4}(\zeta -2)\right] .
\end{equation}

With this choice of the potential for the matter energy density we find
\begin{equation}
r=\frac{4\nu}{3}\theta \left(\theta ^2-\tilde{k}^2\right).
\end{equation}

In the following we assume that the parameter $\zeta $ can take only
positive values, so that $\zeta>0$. The solution of the differential equation (\ref%
{116}) can be obtained as
\begin{align}\label{108}
(1+z)^{3\gamma}=C_1\theta^{\f{\omega}{\zeta}}|\theta-\tilde{k}|^{\f{\tilde{k}-\omega}{\tilde{k}-\zeta}}|\theta+\tilde{k}|^{\f{\tilde{k}+\omega}{\tilde{k}+\zeta}}|\theta-\zeta|^{\f{(\tilde{k}^2-3\zeta^2)(\zeta-\omega)}{\zeta(\tilde{k}^2-\zeta^2)}},
\end{align}
where $C_{1}$ is an arbitrary integration constant that can be obtained by
taking $z=0$ and $\theta \left( 0\right) =\theta _{0}$, where $\theta _{0}$
is the present day value of $\theta $.

For values of $\theta $ so that $\theta\gg\tilde{k}$ and $\theta\gg\zeta $, we
obtain
\begin{equation}
(1+z)^{3\gamma }\approx C_{1}\theta ^{3},
\end{equation}%
and hence we immediately find
\begin{equation}
\theta \approx \theta_0(1+z)^{\gamma }.
\end{equation}%
In this limit the matter energy density can be approximated
as
\begin{equation*}
r\propto \theta ^{3}\propto (1+z)^{3\gamma },
\end{equation*}%
an equation similar to the energy density-redshift relation in the standard general
relativistic cosmology. On the other hand, in the first order approximation,
for the Hubble function, given by
\begin{equation}
h(\theta )=\sqrt{\frac{\nu }{6}}\sqrt{4\theta (\omega -\theta )\left( \theta
^{2}-\tilde{k}^{2}\right) +\left( \frac{v_{0}}{\nu }-2\tilde{k}^{2}\theta
^{2}+\theta ^{4}\right) },
\end{equation}%
we obtain
\begin{equation}
h(\theta )\approx \sqrt{\frac{v_{0}}{6}}-\omega \nu\tilde{k}^{2}\sqrt{\frac{%
2 }{3v_{0}}}\,\theta +O\left( \theta ^{2}\right) ,
\end{equation}%
or, in terms of the redshift,
\begin{equation}
h(z)\approx \sqrt{\frac{v_{0}}{6}}-\omega \nu\tilde{k}^{2}\sqrt{\frac{%
		2 }{3v_{0}}}\,\theta_0(1+z)^{\gamma}.
\end{equation}
For the deceleration parameter we obtain
\begin{equation*}
q(z)\approx -1-\frac{2\gamma \theta _{0}\tilde{k}^{2}\,\omega \,\nu\,(z+1)^{\gamma }}{v_0-2\theta _{0}\tilde{k}%
^{2}\,\omega\,\nu\,(z+1)^{\gamma }}.
\end{equation*}%
Near $z=0$, the deceleration parameter behaves as
\begin{align}
q(z) &\approx-1-\frac{2\gamma \theta _{0}\tilde{k}^{2}\,\omega \,\nu}{v_0-2\theta _{0}\tilde{k}%
	^{2}\,\omega\,\nu} \notag \\
&-\frac{2\gamma^2 \theta _{0}\tilde{k}^{2}\,\omega \,\nu}{(v_0-2\theta _{0}\tilde{k}%
	^{2}\,\omega\,\nu)^2}\,z+O\left( z^{2}\right) .
\end{align}

At $z=0$ we obtain

\begin{equation}
q(0) \approx-1-\frac{2\gamma \theta _{0}\tilde{k}^{2}\,\omega \,\nu}{v_0-2\theta _{0}\tilde{k}%
	^{2}\,\omega\,\nu}
\end{equation}

If the parameters of the model are chosen so that the second term in the
above equation can be neglected with respect to minus one, the Universe
filled with an Einstein type dark energy will reach at the present time a de
Sitter type phase.

\begin{figure*}[t]
\centering
\includegraphics[width=8.5cm]{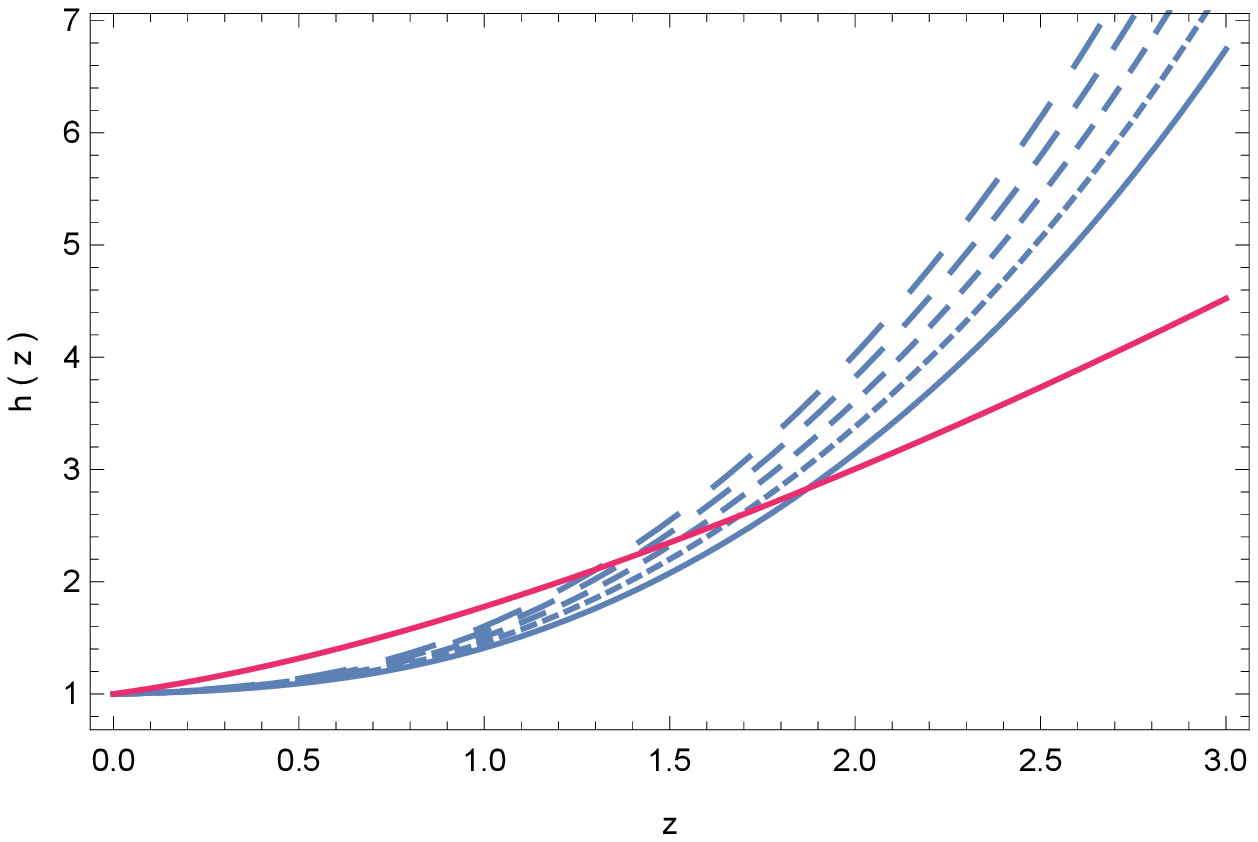} \includegraphics[width=8.5cm]{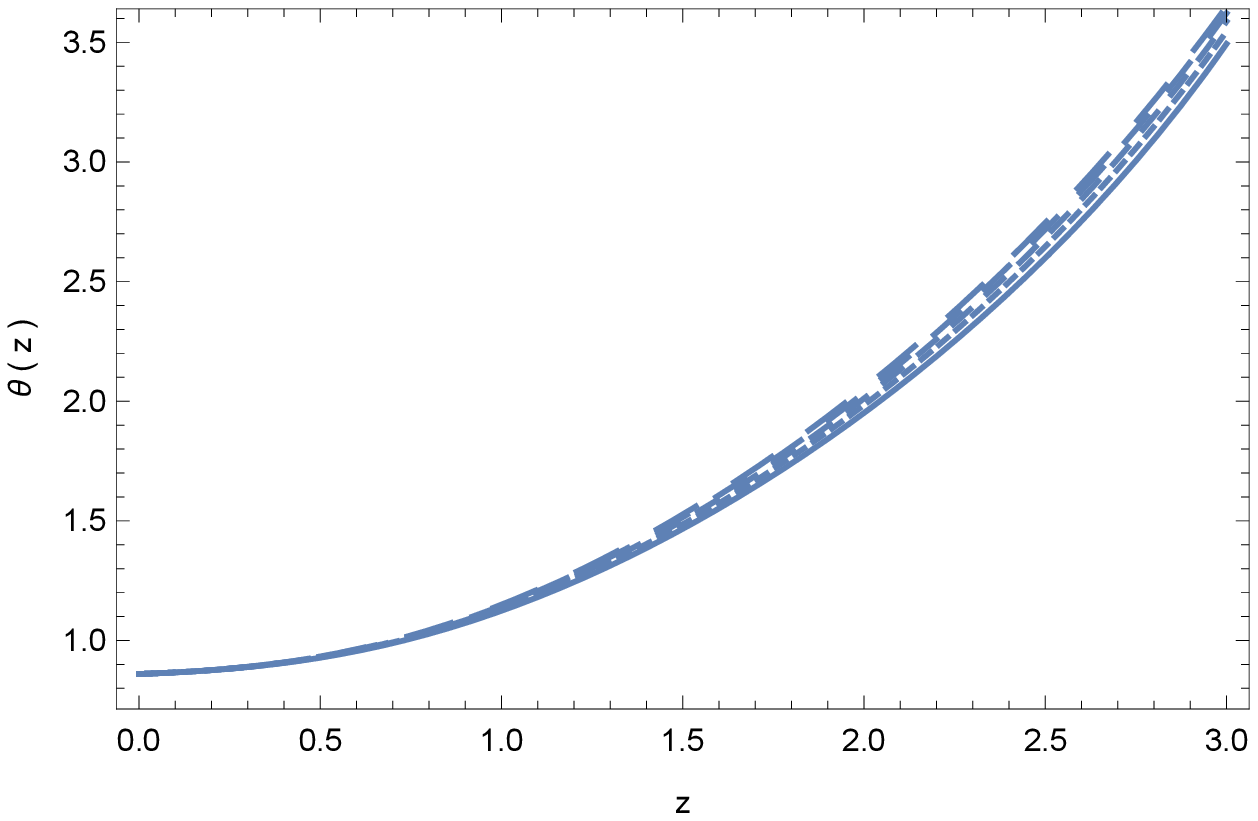}%
\newline
\caption{Variation of the dimensionless Hubble function $h$ (left figure)
and of the dimensionless vector potential $\theta $ of the dark
energy (right figure) as a function of the redshift $z$ in the Einstein dark
energy model for $\gamma =1$, $\tilde{k}=0.85$, $\nu=0.50$,
and for different values of the parameter $\zeta$: $\zeta=10$
(solid curve), $\zeta=11$ (dotted curve), $\zeta=12$
(short dashed curve), $\zeta=13$ (dashed curve), and $\zeta%
=14$ (long dashed curve), respectively. The initial conditions used to
integrate the field equations (\ref{111a}) and (\ref{116})
are $h(0)=1$ and $\theta (0)=0.86$. The red curve in the right figure represents the redshift variation of the dimensionless Hubble function in the standard $\Lambda $CDM cosmological model.}
\label{fig1}
\end{figure*}

\begin{figure*}[t]
\centering
\includegraphics[width=8.5cm]{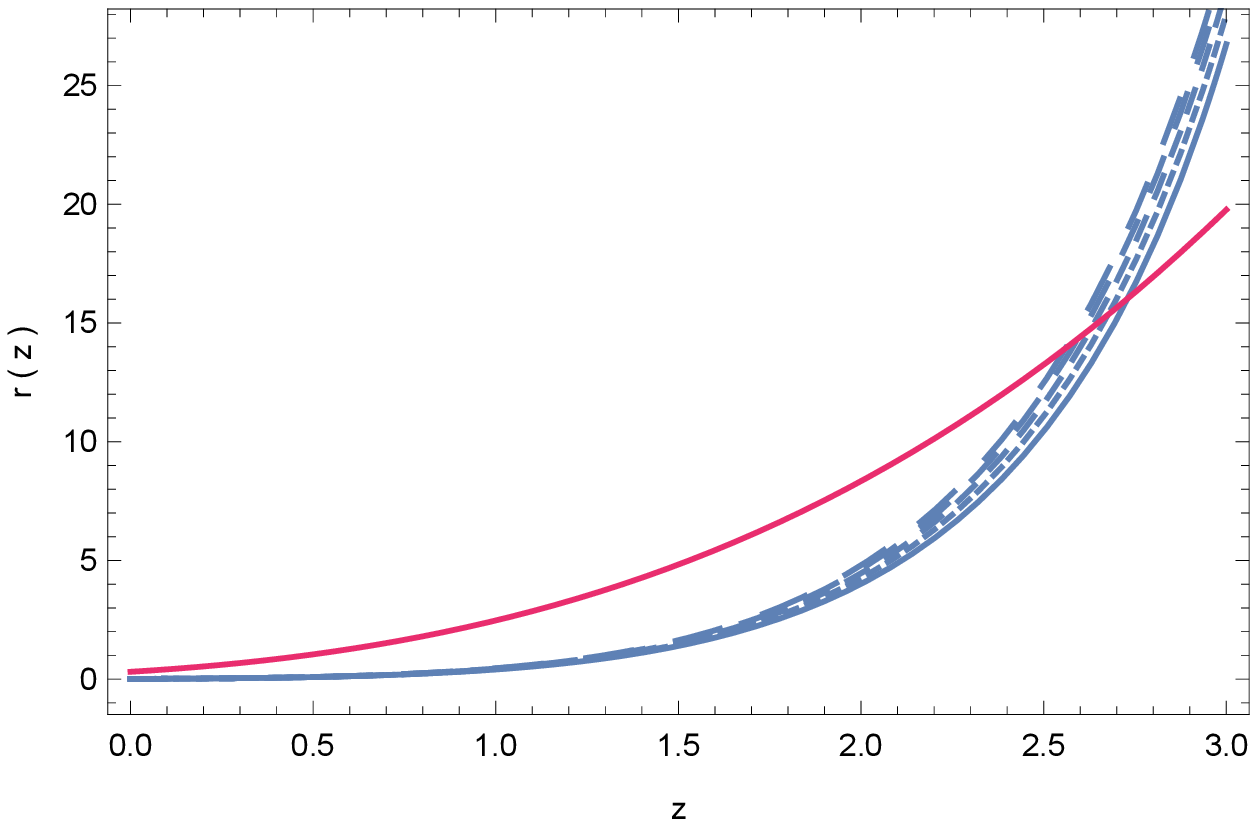} \includegraphics[width=8.5cm]{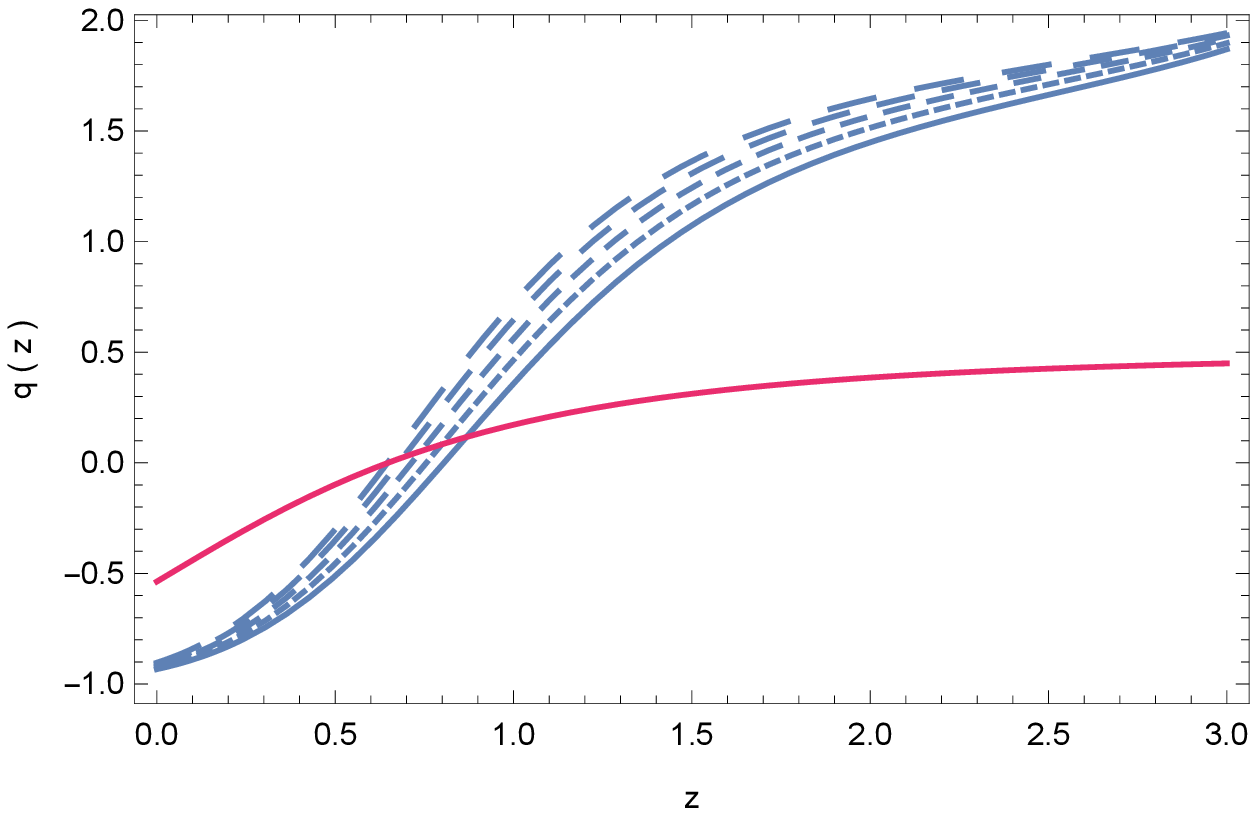}%
\newline
\caption{Variation of the dimensionless matter energy density $r$ (left
figure) and of the deceleration parameter $q $ of the dark energy (right
figure) as a function of the redshift $z$ in the Einstein dark energy model
for $\gamma =1$, $\tilde{k}=0.85$, $\nu=0.5$, and for
different values of the parameter $\zeta$: $\zeta=10$ (solid
curve), $\zeta=11$ (dotted curve), $\protect\zeta=12$ (short
dashed curve), $\protect\zeta=13$ (dashed curve), and $\protect\zeta=14$
(long dashed curve), respectively. The initial conditions used to integrate
the field equations (\protect\ref{111a}) and (\protect\ref{116}) are $h(0)=1$
and $\protect\theta (0)=0.86$. The red curves in the figures represent the redshift variation of the dimensionless matter energy density and of the deceleration parameter in the standard $\Lambda $CDM cosmological model.}
\label{fig2}
\end{figure*}

The variations with respect of the redshift of the Hubble function, of the
dark energy vector potential $\theta$, of the matter energy density and of
the deceleration parameter of the Universe, obtained by numerically
integrating the cosmological evolution equations (\ref{111a}) and (\ref{116}%
) for the case of dust, are presented, for different values of the model
parameter, in Figs.~\ref{fig1} and \ref{fig2}, respectively.

As one can see from the left panel of Fig.~\ref{fig1}, the Hubble function
is an increasing function of the redshift (a decreasing function of the
cosmological time), indicating an expansionary evolution of the Universe
filled with Einstein type dark energy. The evolution of the Hubble function
is strongly dependent on the numerical values of the parameter $\zeta $, and
small variations of this parameter can induce important modifications in the
overall expansion rate of the Universe. In the redshift range $z\in (0,0.2)$
the Hubble function becomes a constant, indicating the presence of a de
Sitter type evolution, which is independent on the numerical values of the
model parameters. The vector field potential $\theta $, depicted in the
right panel of Fig.~\ref{fig1}, is also an increasing function of the
redshift $z$, indicating a time decreasing evolution of the vector field.
The dynamics of $\theta $ is strongly dependent by the considered variations
of the model parameters, and this dependence is particularly important at
high redshifts. The dimensionless matter energy density, represented in the
left panel of Fig.~\ref{fig2}, is a monotonically increasing function of the
redshift, and its evolution at high redshifts is strongly dependent on the
numerical values of the model parameter $\zeta$. But for redshift values in
the range $z\in (0,0.2)$, the variation of $r$ is basically independent of
the variation of $\zeta $. In this model the energy density of the Universe
is generally non-zero at the present time, $r(0)\neq 0$, indicating the
possibility of an accelerated evolution in the presence of (very low
density) dust matter. The deceleration parameter $q$, plotted in the right
panel of Fig.~\ref{fig2}, shows a complex behavior. The evolution of the Universe begins at a redshift $z=3$ from a decelerating state, with $q\approx 2$. The numerical values of $q$ decrease in time, and the deceleration
parameter reaches its marginally accelerating value $q=0$ at around $%
z=z_{cr}\approx 0.65-0.80$. For values of $z<z_{cr}$ the Universe enters in an
accelerating phase with $q<0$, and reaches a state with $q\approx -1$ at the
present time. At high redshifts the evolution of $q$ is dependent on the
numerical values of the model parameter $\zeta $. In the redshift range $%
z\in (0,0.10)$ the evolution of $q$ does not depend essentially on the model parameters.
Moreover, the de Sitter type phase, with $q\approx-1$, reached at $z=0$, is an attractor of the gravitational field equations- no matter the initial conditions of the expansion, the Universe always ends in a state with an exponentially increasing scale factor.

It is also interesting to perform a comparison, at a qualitative level, between the evolution of the Einstein dark energy dominated Universe, and the standard $\Lambda $CDM model. The basic difference between the two models is related to the fact that, as one can see from the right panel in Fig.~\ref{fig2},  in the Einstein dark energy model at the present time the Universe enters in the exact de Sitter phase, with $q=-1$, while in the $\Lambda$CDM model at the present time $q=-0.53$. This means that in the presence of the Einstein type vector dark energy the Universe accelerates more rapidly, with the de Sitter phase with $q\approx -1$ already reached at $z\approx 0.20$. On the other hand, at least for the considered set of model parameters, the predictions for the numerical values of the deceleration parameter are very different. While the $\Lambda$CDM model gives a value of $q\approx 0.1$ at $z=3$, in the present dark energy model the deceleration parameter has values of the order of $q\approx 2$ at $z=3$. However, this value is strongly dependent on the choice of the model parameters $\nu$, $\tilde{k}$, and $\zeta$, and on the initial value/present day value of the vector field $\theta$. Both models predict the presence of cosmological matter at $z=0$. However, due to the much earlier presence of the de Sitter phase, the matter content of the Universe in the Einstein dark energy model is much more diluted than in the standard $\Lambda$CDM model. The evolution of the Hubble function in the $\Lambda$CDM model is different from its evolution in the Einstein dark energy model at both low and high redshifts. At $z=0$ it is not yet a constant, and at high redshifts it has numerical values around two times smaller as compared to the predictions of the Einstein dark energy model.

\section{Discussions and final remarks}

\label{sect6}

In 1919 Einstein proposed an intriguing theory according to which the
gravitational interaction plays an essential role at the level of elementary
particles. The basic assumption in Einstein's theory is that material
particles are held together by the gravitational force, which such
compensates for the presence of the electromagnetic interactions. There are
two major results in Einstein's paper: an explanation of the nature of the
cosmological constant, which naturally appears as a "simple" arbitrary
integration constant, and, more intriguingly, that the energy balance of
material bodies must consist of 25\% gravitational (matter) energy, and 75\%
of electromagnetic type energy.

We believe that the interest for this model may go beyond its historical importance, the possible explanation of the matter-energy composition of the Universe, and of the possibility of solving (at least partially) the cosmological constant problem. The Einstein model is also very interesting, and important, from a general theoretical point of view, since for the first time it did suggest that matter may play a more important role in the framework of gravitational theories as compared to standard general relativity. In the Einstein geometry-matter symmetric field equations matter appears in a mathematically equivalent form with geometry. Hence in this sense Einstein's 1919's theory is a precursor of the present day $f(R,T)$ gravity theories \cite{frt1}. On the other hand Einstein's theory also represents a drastic departure from the standard, and (almost) universally accepted, assumption of the conservation of the matter energy-momentum tensor. Particle creation is also a characteristic feature of quantum field theory in curved
space-times \cite{bookf}, and thus one may interpret non-conservative gravitational models as a first order approximation to quantum gravity phenomenology.  The introduction of a coupling between the quantum fields and the curvature of the space-time, as proposed, for example, in \cite{Kibble},  leads, in the semi-classical approximation, to the non-conservation of the quantum average of the matter energy-momentum tensor operator, so that $\nabla _{\mu}\left<\hat{T}^{\mu \nu}\right>\neq 0$. Thus, theoretical field gravitational models that describe effective particle creation processes can
be interpreted as giving a description at an effective semiclassical level of the quantum processes in a gravitational field.  Therefore a possible physical explanation of the matter production processes in the Einstein dark energy model may be traced back to the semiclassical approximation
of the quantum field theory in a Riemannian curved geometry. On the other hand, if the quantum metric can be decomposed as the sum of the classical and of a fluctuating part, of quantum origin, at the classical level the corresponding Einstein quantum gravitational field equations lead to modified gravity models with a nonminimal coupling between geometry and matter \cite{Liu},  indicating an irreversible transformation of the quantum energy flow of the gravitational field into a matter fluid. Hence Einstein's "theory of elementary particles" may provide some insights into the possibility of the effective description of gravity at quantum level, where the strict distinction between matter and geometry may not exist anymore.

In the present paper we have extended Einstein's elementary particle model
to the scale of the biggest possible particle, the Universe. We have
substituted the "simple" electromagnetic force by a vector type dark energy,
and we have generalized the Einstein model by assuming that the vector dark
energy is massive, has self-interaction, described by a corresponding
potential, and couples with the matter current. Under these assumptions we
have formulated the variational principle from which the field equations of
the Einstein dark energy model can be obtained. The Einstein field equations
from 1919 \cite{Ein2} can also be obtained as a limiting case from this variational
principle. The initial Einstein theory is non-conservative, and
this feature was automatically transferred to its generalization. Non-conservation of matter can be described naturally in
the framework of the thermodynamics of open systems as describing matter and
entropy creation through the transfer of gravitational energy.

In the present paper we have also investigated in detail the cosmological
properties of the Einstein cosmological models. The vector field
self-interaction potential was assumed to be of a simple polynomial form,
constructed by analogy with the Higgs potential \cite{Aad}, which plays a
fundamental role in theoretical particle physics as describing the
generation of mass of the quantum elementary particles.

In the framework of the Einstein dark energy model we have investigated two
cosmological evolution scenarios, corresponding to matter conservation, and
non-conservation, respectively. In the non-conservative case, in the large time (small
redshift) limit, the Universe enters in an accelerating stage, with the
exponential de Sitter solution acting as an attractor for these cosmological
models. The deceleration parameter reaches the marginal $q=0$ value at a
redshift of the order of $z\approx 0.65-0.80$. However, the time
evolution of the deceleration parameter strongly depends on the numerical
values of the model parameters, describing the mass of the dark energy
field, the coupling between dark energy and matter current, and of the
functional form of the self-interaction potential $V$. In order to obtain a
better understanding on the numerical values of these parameters the fitting
of the model with the cosmological observational data should be performed. This in depth comparison of the theoretical predictions with the observational data can give the answer to the question if Einstein did really  predict almost 100 years ago the correct "chemical and
energy composition" of the Universe.

Hopefully the Einstein dark energy model introduced in the present paper could
give some new insights into the complex problem of the evolutionary
dynamics, composition and structure of the Universe, from its birth to the
latest stages of evolution.

\section*{Acknowledgments}

We would like to thank the anonymous referee for comments and suggestions that helped us to significantly improve our manuscript.

\end{document}